# Temperature activation of electrical conductivity and self-healing in silicone rubber/graphite nanoplatelets composites

S. Bittolo Bon[a], L. Valentini[a*]

[a]Civil and Environmental Engineering Department and INSTM Research Unit, University of Perugia, Strada di Pentima, 4, 05100 Terni, Italy; E-mail: luca.valentini@unipg.it

With the request of higher performance in automotive products, sealing components and materials resisting to severe conditions, the performance requirements for silicones are becoming ever more diverse and sophisticated. In this article we present silicone rubber composite designed for multifunctional applications. The development of silicone rubber nanocomposite with electrical conductivity activated by temperature and self-healing capabilities has been reported. Silicon rubber (SR) filled with graphite nanoplatelets (GNPs) was synthesized via a liquid mixing method. The electrical conductivity of the SR/GNPs composite showed a transition from insulating to conducting state with increasing the temperature. The SR/GNPs composite can be healed by simple thermal annealing. These results, together with the wide applicability of the silicone rubber material, indicate that SR/GNPs composites can act as a promising self-healing material and find wide applications by developing new high-performance silicone rubber components for seals, hoses and automotive sector.

1. Introduction

Rubber materials are commonly considered the workhorse of the industrial and automotive rubber products industries [1-3]. As a significant candidate of them, the silicone rubber (SR) offers a unique combination of chemical and mechanical properties that organic elastomers cannot match. These properties make silicone rubber the material of choice for applications ranging from bakeware to turbocharger hoses due to its good biocompatibility, thermal stability and especially excellent elasticity [4,5]. Moreover, such soft and flexible materials have attracted attention due to their potential applications in advanced strain sensors for the fabrication of artificial skins [6-8].

Nevertheless, elastomeric applications are also susceptible to mechanical and chemical damage (e.g. scratches, cuts and punctures). Such damages result in the loss of the originally intended functions of the elastomers leading to spillage, contamination, safety hazards or just lost of performance. Such damages are particularly problematic when the elastomers are used as seals, hoses and coatings. After the occurrence of the first cracks or surface damage, the material is especially susceptible to further damage. For this reason, seals, hoses or coatings have to be frequently checked and mended or even replaced with the subsequent cost. Healable polymers offer an alternative to the damage-and-discard cycle and represent a first step in the development of polymeric materials that have much greater lifespans than currently available.

For decades, scientists and engineers have paid great attention to developing self-healing polymeric materials to improve the safety and lifetime [9]; the most used approach consists in the utilization of microcapsules [10-12] which store the healing agents into the polymer matrix. Once a crack is formed, the healing agent is released into the crack due to the capillary effect, and then re-bond the crack. In this way the material can be healed only once at the same location because of exhaustion of the healing agents.

Zheng et al. [13] have demonstrated that silicone rubber that has been cut in half can completely repair itself through heat-activated reversible bonding.

Graphite nanoplatelets, have recently attracted attention as a viable and inexpensive material [14-16] that can be used in many engineering applications, given the excellent in-plane mechanical, structural, thermal, and electrical properties of graphite.

Because thermal conductivity of elastomers is very low, the heat build-up is harmful to elastomers, because elastomers are susceptible to thermal degradation [17,18]. Composites with carbon-based fillers showed thermal stability, light weight, and high thermal conductivity [19,20].

These properties prompt us to believe that integrating graphite nanoplatelets (GNPs) with SR might generate a novel self-healing materials, which deals multifunctional properties such as self-healing by thermal annealing and temperature activation of the electrical conductivity. To our knowledge, this is the first example of self-healable SR composite with graphite nanoplatelets that switches to conductive state with increasing its temperature. Our material paves potential applications in sealing and extreme thermal environments.

2. Experimental section

Liquid rubber (GLS-50 purchased from PROCHIMA®) was used for casting with a cold cure by poly-condensation. Before using, we add to the rubber 5% of T30 catalyst (purchased from PROCHIMA®). The complete vulcanization takes 18-20 hours at room temperature. To accelerate this process, the blend was put in a warm place (30 C°), but the reaction was too fast and did not allow the escape of air bubbles.

GNPs were purchased from Cheaptubes. GNPs were dispersed in liquid silicone rubber (1 %wt.) through the utilization of a Dispermat (500 rpm for 1hr) to facilitate the dispersion of GNPs. Then the catalyst was added. The liquid composite was deposited onto a silicone mould and the vulcanization was obtained in 18-20 hrs at room temperature.

The melting behaviours of the samples were tested by differential scanning calorimeter (DSC) using a TA Q200 DSC analyser under nitrogen atmosphere. Samples were heated from -80°C to 150°C, cooled to -80°C, and heated to 150°C again. In all heating and cooling cases, the rate was set at 10°Cmin$^{-1}$.Thermogravimetric analysis (TGA) were performed in nitrogen with a TG/DTA Extar 6300 at a heating rate of 10°C min$^{-1}$.

The electrical characterization of both neat SR and SR/GNPs composite was performed, by using a computer controlled Keithley4200 source measure unit. The electrical current was recorded by biasing the samples at 50V and 100V at different temperatures.

The samples were cut into strips of ~ 100 mm × 10 mm × 2,5 mm. Before healing, a 3 mm cut was made in the middle of the sample along the strip traverse direction, and then the cut sample was healed by thermal annealing at 250°C for 2hrs. The mechanical properties were measured by an universal tensile testing machine (Lloyd Instr. LR30K) with a 250 N cell at room temperature. The extension rate was 50 mm*min$^{-1}$ and the gauge length was 50 mm.

3. Results and discussion

The thermal properties of neat SR and SR/GNPs composites were investigated. Figures 1(a) and 1(b) show the TGA curves and the corresponding differential thermogravimetric (DTG) analysis curves. Figure 1(b) showed that both samples decomposed with a one-step process, which meant the GNPs did not break the network of the SR. The incorporation of the GNPS did not change the onset temperature of the composites compared to that of the neat SR.

DSC analysis for the SR and the SR/GNR nanocomposites was then performed and the results were shown in figure 2. It was known from the literature [21] that the glass transition temperature of the SR was lower than -100°C, so the endothermic peaks observed in figure 2 correspond to the melting temperature. It was found that the presence of the nanofiller did not affect the melting temperature of the neat SR matrix.

Silcone rubber has high insulation resistance of 1TΩ*m – 100 TΩ*m, and its insulating properties are stable over a wide range of temperature as confirmed by the data reported in figure 3(a) where no change of the electrical conductivity was detected biasing the sample with 50V up to 250°C.

Carbon nanomaterials are ideal conductive fillers that are able to enhance both the electrical and the mechanical properties of the elastomers [22]. The formation of a conductive network by the fillers into the matrix infer the conductivity to the polymer matrix. The conductive network is thus formed due to the contact of the laminar GNPs with each other. When GNPs are more, a large number of graphene would be aggregated which prevent the formation of conductive path, resulting in poor conductivity. Thus the electric conduction occurs based on the "tunneling effect", which postulates that the GNPs cannot make a perfect network in the insulating matrix. In our case the GNPs content did not allow to reach the electrical percolation threshold thus an insulating behaviour was observed at room temperature for the SR composite when a bias of 50V was applied across it (inset of figure 3(b)). As a result, the electrons cannot directly move between the fillers and instead, they trapped in the potential barrier leading to a poor conductivity.

Figure 2(b) shows that raising the temperature of the composite up to 250° we observed a transition to a conducting state being this effect more evident by increasing the bias voltage. When the temperature increases the probability of the electronic transition raises, as a result an increase of the conductivity of the composite was observed. This single mechanism could be attributed to grain site or due to the mobility of free charges (polaron or free ions) induced by the increment of temperature [23].

The mechanical performance of the pristine SR and the SR/GNPs nanocomposites was then tested in terms of typical strain–stress behaviour, and the results were shown in figure 4(a) and Table 1. As shown, the mechanical performance of the SR/GNPs composite had an improvement of the elastic modulus compared to that of the pristine SR. The reduction of the elongation at break as well as that of the tensile strength is consistent with the increase of the elasticity of the composite.

Optical images were used to monitor the healing of the SR/GNPs composite. The optical images for the cracked sample before and after thermal annealing are shown in figure 4(b). From the comparison of the two images it is evident that the crack in the SR/GNPs sample disappeared after healing. As reported in table 1 the healing efficiencies of the mechanical properties of the SR/GNPs composite are 94% and 90% for the elastic modulus and the tensile strength, respectively. The kinetic of reaction in this system requires activation of the damage at use temperature with heat. Thus these high yields can be explained considering the excellent thermal conductivity of graphite nanoplatelets that transfer the heat into the SR matrix. Thus, in the healing process, graphite nanoplatelets work as a nanoscale heater and transfer unit to generate the required energy and then transport the energy to the matrix efficiently.

4. Conclusions

In summary, novel multifunctional properties of silicone rubber composite system have been reported. Graphite nanoplatelets enhanced the thermal transport leading to a self-healable SR/GNPs composite with high yields. Thus it was demonstrated that such silicone composite can be healed by simple thermal annealing. Furthermore, such composite remains electrically insulating till the temperature activates the electrical conductivity. The remarkable and variable properties of such silicones as well as their broad use as materials suggest that these findings will be broadly applicable.

Figures

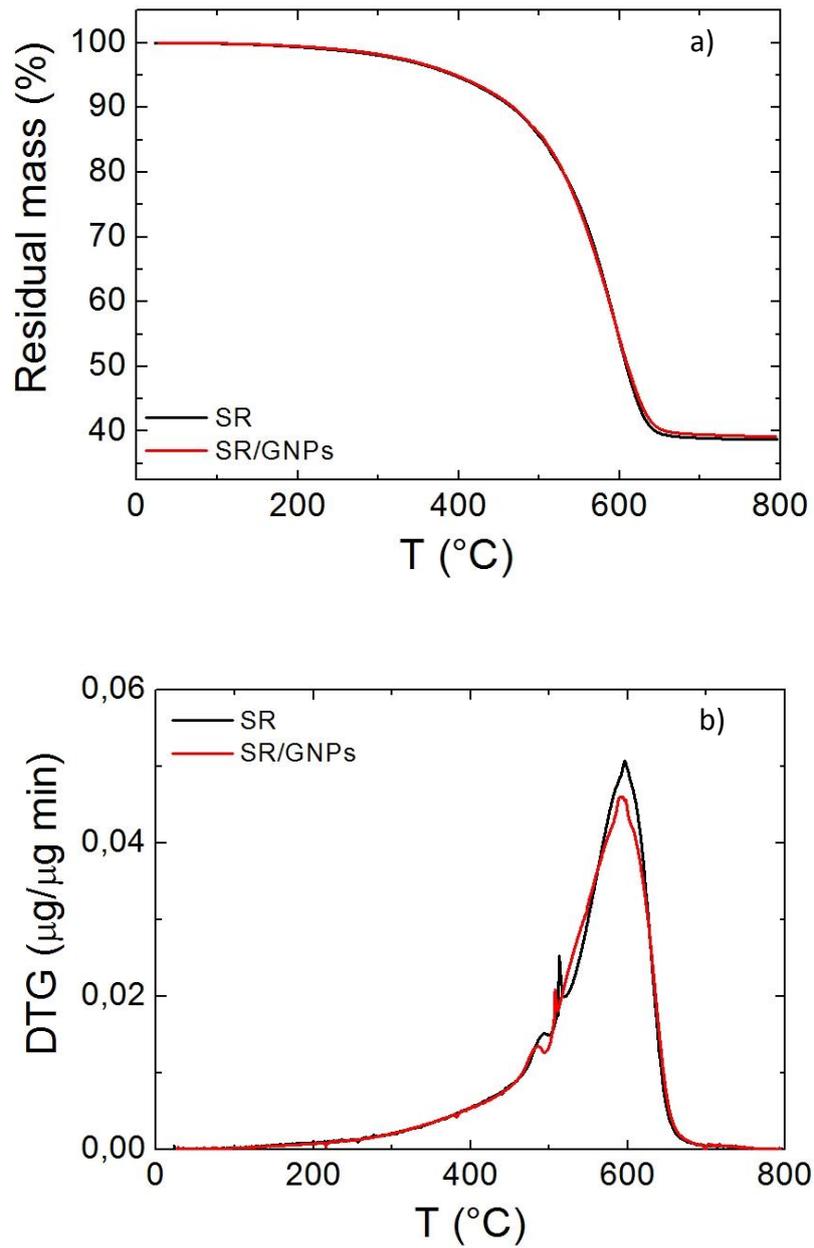

Figure 1. (a) TGA curves and corresponding (b) DTG curves of the pristine SR and the SR/GNPs nanocomposite.

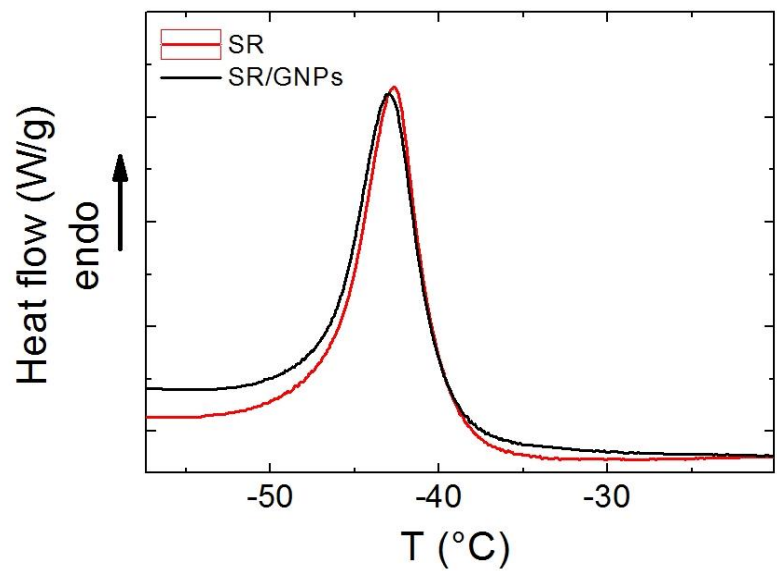

Figure 2. DSC curves of the pristine SR and the SR/GNPs nanocomposite.

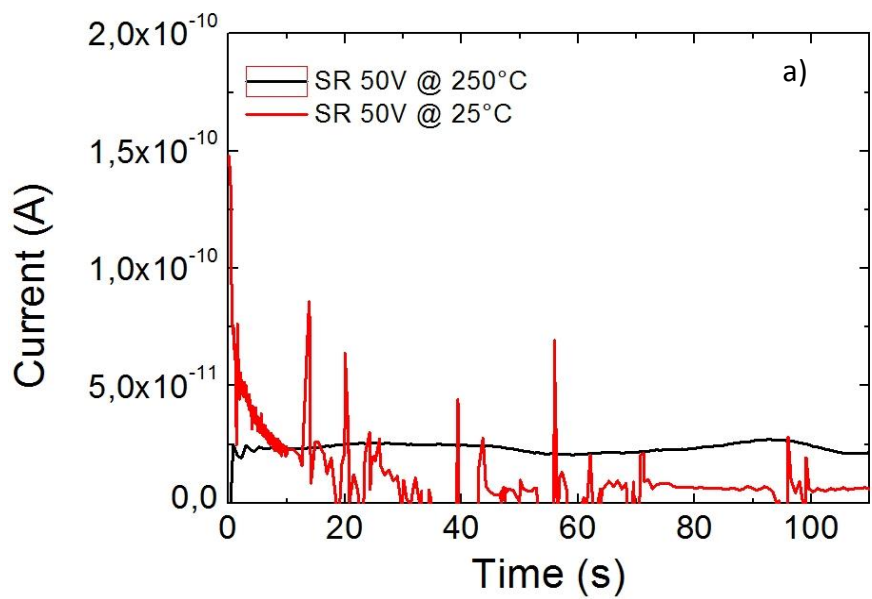

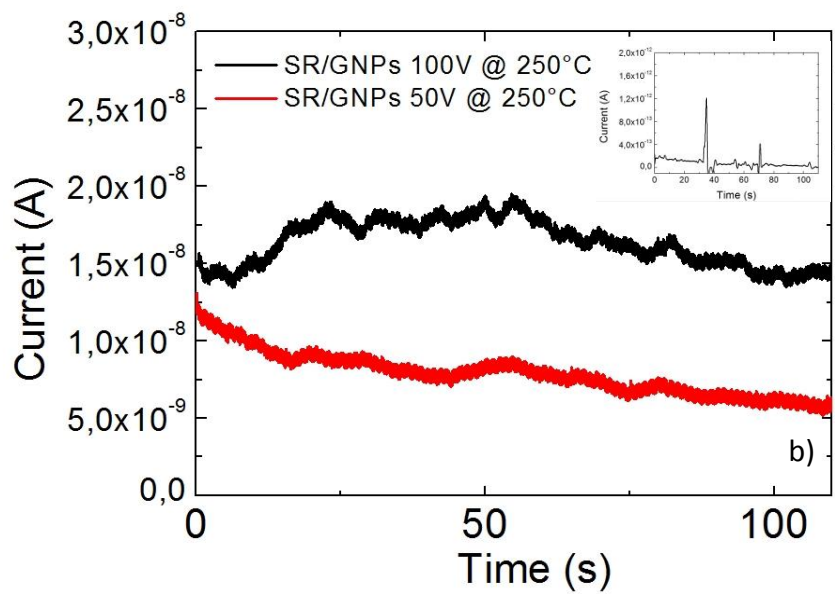

Figure 3. Current–time data of (a) SR and (b) SR/GNPs composite recorded at different bias voltages and temperatures. The inset of panel (b) shows the current-time data of SR/GNPs composite biased with 50V at room temperature.

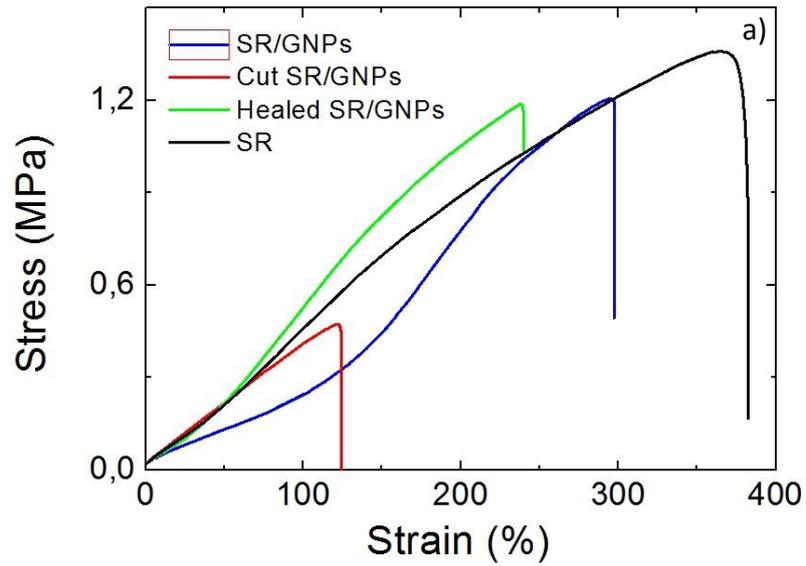

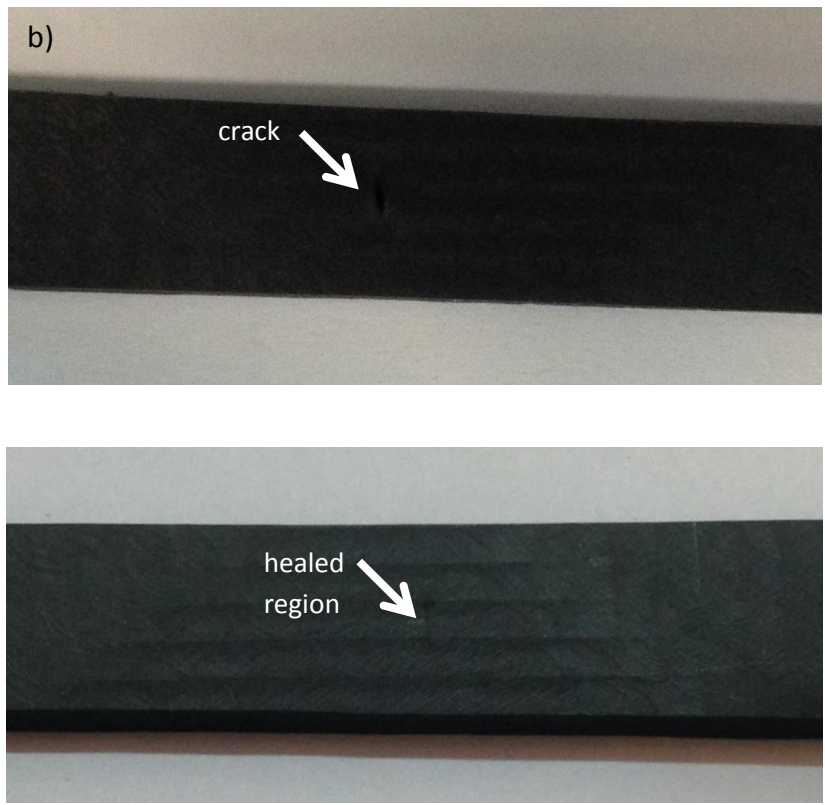

Figure 4. (a) Stress–strain curves of neat SR, SR/GNPs composite, cut composite and healed composite. (b) Optical images of the SR/GNPs sample before and after healing by thermal annealing.

Table 1. Tensile properties of neat SR, SR/GNPs composite, cut composite and healed composite.

| Sample | Young Modulus (MPa) | Tensile Strength (MPa) | Elongation at break (%) |
|---|---|---|---|
| SR | 0,51±0,11 | 1,35±0.10 | 365,22 |
| SR/GNPs | 0,69±0.14 | 1,20±0.36 | 297,26 |
| Cut SR/GNPs | 0,44±0.10 | 0,47±0.03 | 124,06 |
| Healed SR/GNPs | 0,65±0.05 | 1.18±0.10 | 239,51 |